\documentclass[a4paper,11pt]{article}
\usepackage{jinstpub}

\usepackage{longtable}
\usepackage{multirow} 
\usepackage{lscape}

\usepackage{lineno}

\title{Radiopurity assessment of the energy readout for the NEXT double beta decay experiment}

\author[a,b,1]{S.~Cebri\'an,\note{Corresponding author.}}
\author[c]{J.~P\'{e}rez,}
\author[b]{I.~Bandac,}
\author[d]{L.~Labarga,}
\author[c]{V.~\'Alvarez,}
\author[e]{C.D.R.~Azevedo,}
\author[c]{J.M.~Benlloch-Rodr\'{i}guez,}
\author[f]{F.I.G.M.~Borges,}
\author[c]{A.~Botas,}
\author[c]{S.~C\'arcel,}
\author[c]{J.V.~Carri\'on,}
\author[f]{C.A.N.~Conde,}
\author[c]{J.~D\'iaz,}
\author[g]{M.~Diesburg,}
\author[f]{J.~Escada,}
\author[h]{R.~Esteve,}
\author[c]{R.~Felkai,}
\author[i]{L.M.P.~Fernandes,}
\author[c]{P.~Ferrario,}
\author[e]{A.L.~Ferreira,}
\author[i]{E.D.C.~Freitas,}
\author[j]{A.~Goldschmidt,}
\author[c,2]{J.J.~G\'omez-Cadenas,\note{\label{note1}NEXT Co-spokesperson.}}
\author[k]{D.~Gonz\'alez-D\'iaz,}
\author[l]{R.M.~Guti\'errez,}
\author[m]{J.~Hauptman,}
\author[i]{C.A.O.~Henriques,}
\author[l]{A.I.~Hernandez,}
\author[k]{J.A.~Hernando~Morata,}
\author[h]{V.~Herrero,}
\author[n]{B.J.P.~Jones,}
\author[c]{A.~Laing,}
\author[g]{P.~Lebrun,}
\author[c]{I.~Liubarsky,}
\author[c]{N.~L\'opez-March,}
\author[l]{M.~Losada,}
\author[c,3]{J.~Mart\'in-Albo,\note{Now at University of Oxford, United Kingdom.}}
\author[k]{G.~Mart\'inez-Lema,}
\author[c]{A.~Mart\'inez,}
\author[n]{A.D.~McDonald,}
\author[n]{F.~Monrabal,}
\author[i]{C.M.B.~Monteiro,}
\author[h]{F.J.~Mora,}
\author[e]{L.M.~Moutinho,}
\author[c]{J.~Mu\~noz Vidal,}
\author[c]{M.~Musti,}
\author[c]{M.~Nebot-Guinot,}
\author[c]{P.~Novella,}
\author[n,2]{D.R.~Nygren}
\author[c]{B.~Palmeiro,}
\author[g]{A.~Para,}
\author[c]{M.~Querol,}
\author[c]{J.~Renner,}
\author[o]{L.~Ripoll,}
\author[c]{J.~Rodr\'iguez,}
\author[n]{L.~Rogers,}
\author[f]{F.P.~Santos,}
\author[i]{J.M.F.~dos~Santos,}
\author[c]{A.~Sim\'on,}
\author[p,4]{C.~Sofka,\note{Now at University of Texas at Austin, USA.}}
\author[c]{M.~Sorel,}
\author[p]{T.~Stiegler,}
\author[h]{J.F.~Toledo,}
\author[c]{J.~Torrent,}
\author[q]{Z.~Tsamalaidze,}
\author[e]{J.F.C.A.~Veloso,}
\author[a,b]{J.A.~Villar,}
\author[p]{R.~Webb,}
\author[p,5]{J.T.~White,\note{Deceased.}}
\author[c]{and N.~Yahlali}
\affiliation[a]{
Laboratorio de F\'isica Nuclear y Astropart\'iculas, Universidad de Zaragoza\\
Calle Pedro Cerbuna, 12, 50009 Zaragoza, Spain}
\affiliation[b]{ 
Laboratorio Subterr\'aneo de Canfranc\\ Paseo de
los Ayerbe s/n, 22880 Canfranc Estaci\'on, Huesca, Spain}
\affiliation[c]{
Instituto de F\'isica Corpuscular (IFIC), CSIC \& Universitat de Val\`encia\\
Calle Catedr\'atico Jos\'e Beltr\'an, 2, 46980 Paterna, Valencia, Spain}
\affiliation[d]{
Departamento de F\'isica Te\'orica, Universidad Aut\'onoma de Madrid\\
Campus de Cantoblanco, 28049 Madrid, Spain}
\affiliation[e]{
Institute of Nanostructures, Nanomodelling and Nanofabrication (i3N), Universidade de Aveiro\\
Campus de Santiago, 3810-193 Aveiro, Portugal}
\affiliation[f]{
LIP, Department of Physics, University of Coimbra\\
P-3004 516 Coimbra, Portugal}
\affiliation[g]{
Fermi National Accelerator Laboratory\\
Batavia, Illinois 60510, USA}
\affiliation[h]{
Instituto de Instrumentaci\'on para Imagen Molecular (I3M), Centro Mixto CSIC -— Universitat Polit\`ecnica de Val\`encia\\
Camino de Vera s/n, 46022 Valencia, Spain}
\affiliation[i]{
LIBPhys, Physics Department, University of Coimbra\\
Rua Larga, 3004-516 Coimbra, Portugal}
\affiliation[j]{
Lawrence Berkeley National Laboratory (LBNL)\\
1 Cyclotron Road, Berkeley, California 94720, USA}
\affiliation[k]{
Instituto Gallego de F\'isica de Altas Energ\'ias, Univ.\ de Santiago de Compostela\\
Campus sur, R\'ua Xos\'e Mar\'ia Su\'arez N\'u\~nez, s/n, 15782 Santiago de Compostela, Spain}
\affiliation[l]
{Centro de Investigaci\'on en Ciencias B\'asicas y Aplicadas, Universidad Antonio Nari\~no\\
Sede Circunvalar, Carretera 3 Este No.\ 47 A-15, Bogot\'a, Colombia}
\affiliation[m]{
Department of Physics and Astronomy, Iowa State University\\
12 Physics Hall, Ames, Iowa 50011-3160, USA}
\affiliation[n]{
Department of Physics, University of Texas at Arlington\\
Arlington, Texas 76019, USA}
\affiliation[o]{
Escola Polit\`ecnica Superior, Universitat de Girona\\
Av.~Montilivi, s/n, 17071 Girona, Spain}
\affiliation[p]{
Department of Physics and Astronomy, Texas A\&M University\\
College Station, Texas 77843-4242, USA}
\affiliation[q]{
Joint Institute for Nuclear Research (JINR)\\
Joliot-Curie 6, 141980 Dubna, Russia}

\emailAdd{scebrian@unizar.es}

\abstract{The ``Neutrino Experiment with a Xenon Time-Projection
Chamber'' (NEXT) experiment intends to investigate the neutrinoless
double beta decay of $^{136}$Xe, and therefore requires a severe
suppression of potential backgrounds. An extensive material
screening and selection process was undertaken to quantify the
radioactivity of the materials used in the experiment. Separate
energy and tracking readout planes using different sensors allow us
to combine the measurement of the topological signature of the event
for background discrimination with the energy resolution
optimization. The design of radiopure readout planes, in direct
contact with the gas detector medium, was especially challenging
since the required components typically have activities too large
for experiments demanding ultra-low background conditions. After
studying the tracking plane, here the radiopurity control of the
energy plane is presented, mainly based on gamma-ray spectroscopy
using ultra-low background germanium detectors at the Laboratorio
Subterráneo de Canfranc (Spain). All the available units of the
selected model of photomultiplier have been screened together with
most of the components for the bases, enclosures and windows.
According to these results for the activity of the relevant
radioisotopes, the selected components of the energy plane would
give a contribution to the overall background level in the region of
interest of at most $2.4\times10^{-4}$~counts
keV$^{-1}$~kg$^{-1}$~y$^{-1}$, satisfying the sensitivity
requirements of the NEXT experiment.}


\begin{document}
\maketitle
\flushbottom

\section{Introduction}

\begin{figure}
\centering
  \includegraphics[width=.8\textwidth]{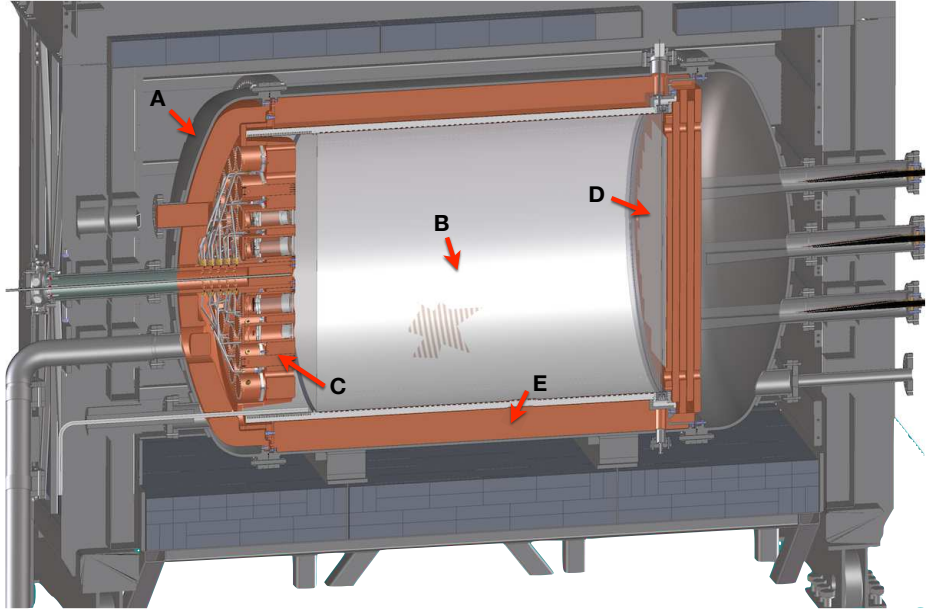}
  \caption{Cross-section view of the NEXT-100 detector inside its lead castle shield. A stainless steel pressure vessel (A) houses the electric-field cage (B) and the two sensor planes (energy plane, C; tracking plane, D) located at
opposite ends of the chamber. The active volume is shielded from
external radiation by at least 12~cm of copper (E) in all
directions.}
  \label{next100}
\end{figure}

Double beta decay is a very active research topic in Neutrino
Physics. The observation of the neutrinoless mode, as a peak at the
transition energy, would give unique information on the neutrino
nature and mass (see for instance \cite{dbdrefs1}-\cite{dbdexp}).
Since it is a very rare process, an ultra-low background level in
the region where the signal is expected to appear is a must for this
kind of experiment. NEXT (``\underline{N}eutrino
\underline{E}xperiment with a \underline{X}enon
\underline{T}ime-Projection Chamber'') \cite{next} aims to search
this in $^{136}$Xe at the Canfranc Underground Laboratory
(Laboratorio Subterráneo de Canfranc, LSC) \cite{lsc}, located in
the Spanish Pyrenees, with a source mass of $\sim$100 kg (NEXT-100
phase). The NEXT-100 detector is designed as an electroluminescent
high-pressure xenon gas Time Projection Chamber (TPC) with two
important features: very good energy resolution (better than 1\%
FWHM at the transition energy of $^{136}$Xe,
Q$_{\beta\beta}=$2.458~MeV) and topological reconstruction for the
discrimination of signal and background events. As sketched in
figure~\ref{next100}, light from the Xe electroluminescence
generated at the anode is recorded both in the photosensor plane
right behind it for tracking and in the plane behind the transparent
cathode at the opposite side of the pressure vessel for a precise
energy measurement. The separate energy and tracking readout planes
use different sensors. Photomultiplier tubes (PMTs) are used for
calorimetry, and for determining the start of the event thanks to
the detection of the primary scintillation. Silicon photomultipliers
(SiPMs) are used for tracking. After successful work on prototypes
\cite{berkeley}-\cite{nextrecognition}, the NEW (NEXT-WHITE)
detector \footnote{The name honours the memory of the late Professor
James White, key scientist of the NEXT project.} is fully
operational at LSC. It is the first phase of the NEXT detector to
operate underground; the NEW apparatus is a downscale 1:20 in mass
of NEXT-100. A low background run using depleted Xe is foreseen
during 2017; these data will be essential to validate the background
model of NEXT-100, based on radiopurity measurements of all the
relevant components, like the ones presented here.

As shown in \cite{nextsensitivity}, an excellent sensitivity is
expected for NEXT-100. For a background rate of $4\times10^{-4}$
counts keV$^{-1}$ kg$^{-1}$ y$^{-1}$ in the energy region of
interest, the experiment is sensitive to a neutrinoless decay half
life up to 6$\times$10$^{25}$ years after running for 3 effective
years. The required background level is achievable thanks to passive
shieldings, background discrimination techniques based on charged
particle tracking \cite{nextrecognition,neuralnetwork} and a
thorough material radiopurity control. The most dangerous background
sources are $^{208}$Tl and $^{214}$Bi, isotopes of the progeny of
$^{232}$Th and $^{238}$U, because of their ability to generate a
signal-like track in the fiducial volume with energy around
Q$_{\beta\beta}$.

A material screening and selection process for NEXT components has
been underway for several years. Determination of the activity
levels is based on gamma-ray spectroscopy using ultra-low background
germanium detectors at LSC and also on other techniques like Glow
Discharge Mass Spectrometry (GDMS) and Inductively Coupled Plasma
Mass Spectrometry (ICPMS). Materials to be used in the shielding,
pressure vessel, electroluminescence and high voltage components,
and energy and tracking readout planes have been measured and
results have been presented in \cite{jinstrp}-\cite{lrtrp}. These
results are the input for the construction of a precise background
model of the NEXT experiment based on Monte Carlo simulations
\cite{nextsensitivity}. The design of radiopure readout planes is
complicated by the fact that sensors, printed circuit boards and
electronic components, involving typically different composite
materials, show in many cases activity levels too large to be used
in experiments demanding ultra-low background conditions (see for
instance \cite{heusser}-\cite{radiopurityorg}). Exhaustive screening
programs specifically for both the tracking and energy planes were
undertaken; the former was presented in \cite{trackingrp} and the
latter is described here.

\begin{figure}
\centering
  \includegraphics[width=.8\textwidth]{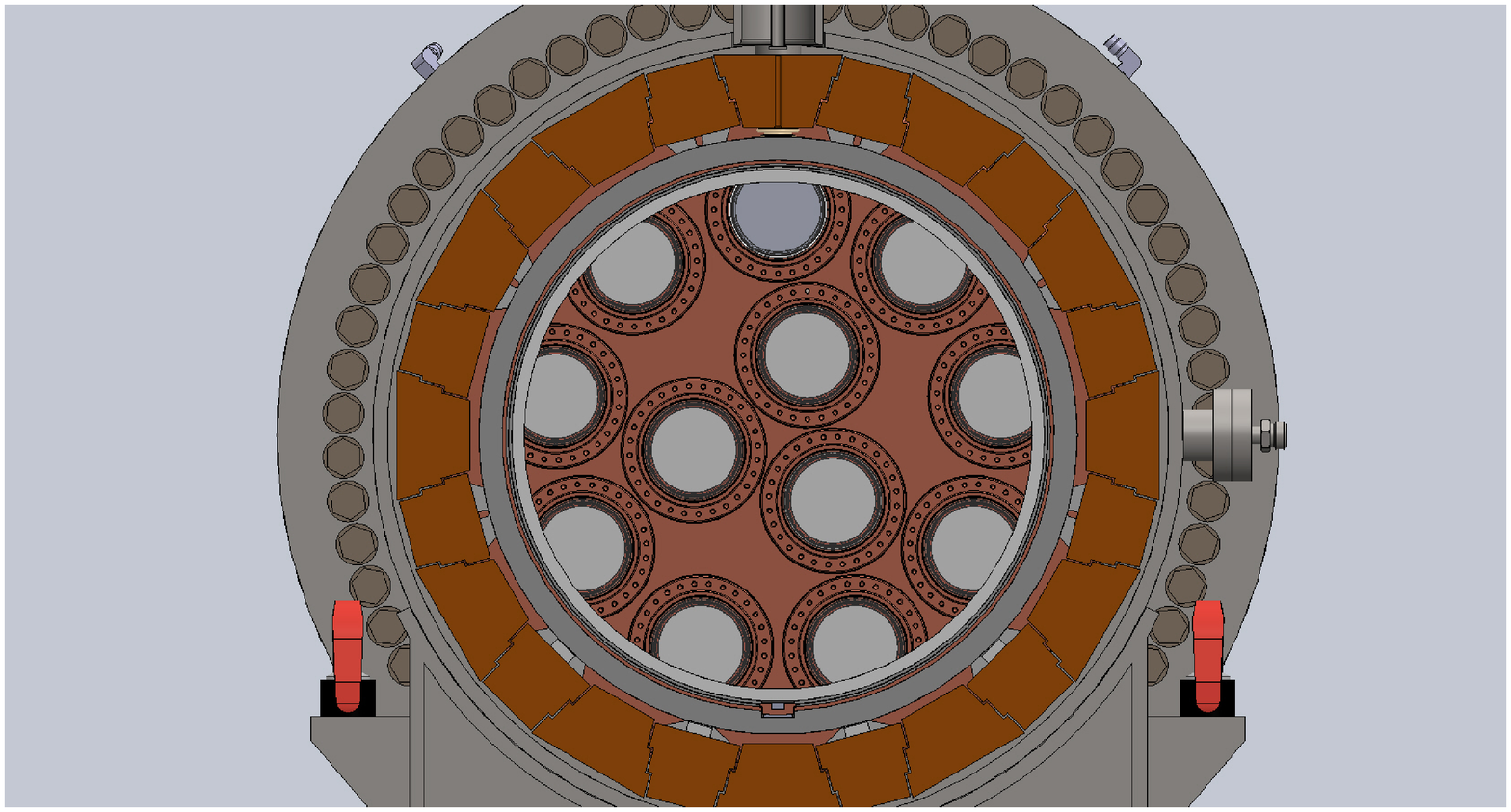} 
  \includegraphics[width=.8\textwidth]{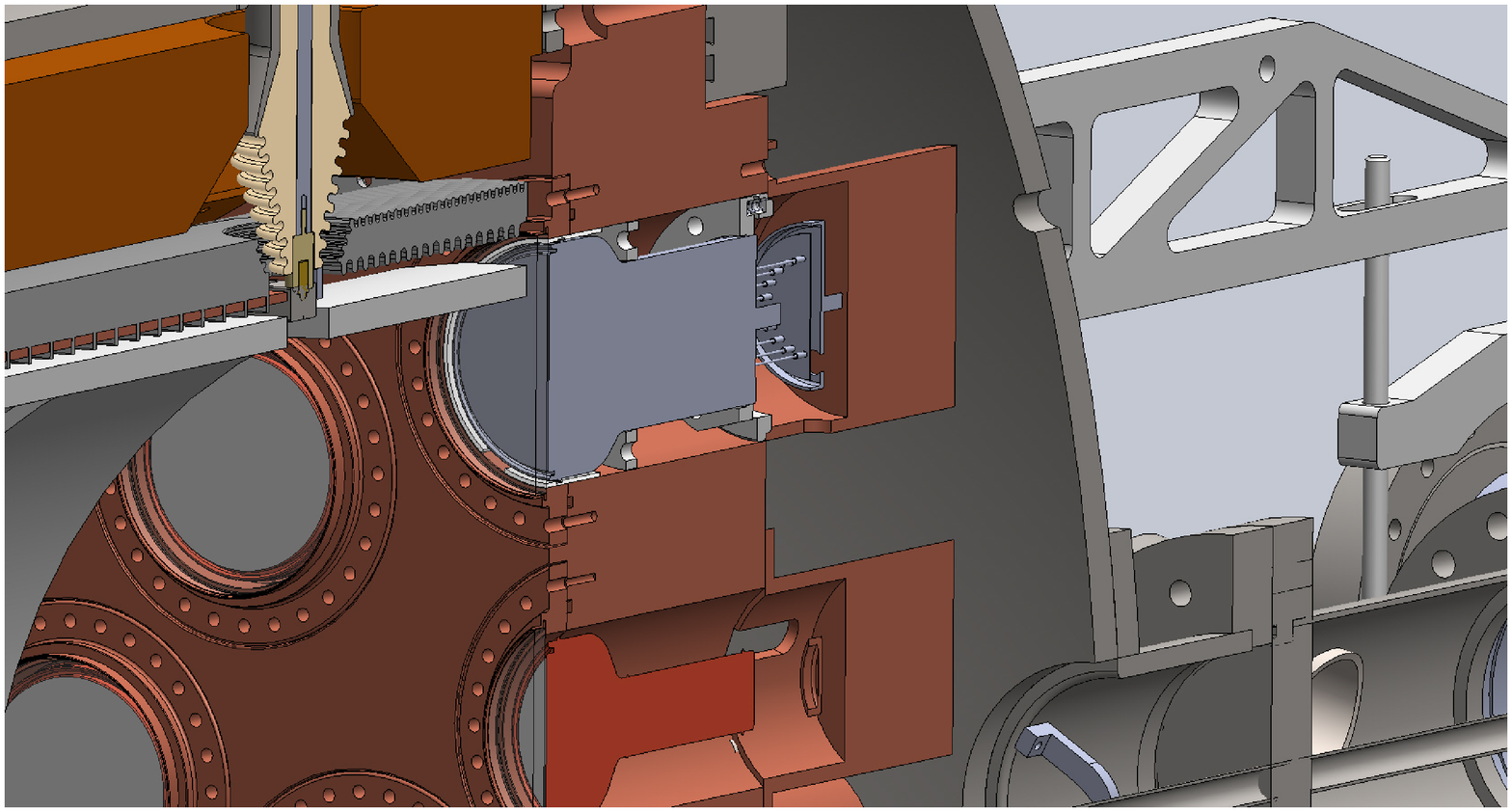}
  \caption{Design of the energy plane of the NEW detector. Top: View of the 12 PMTs and windows, including each one 28 bolts per flange, and the copper plate.
  Bottom: Zoomed image showing the PMT body (in red), the pins connecting the PMT body with the base, and the copper caps behind the PMT base.}
  \label{drawings}
\end{figure}

Figure~\ref{drawings} shows some drawings of the energy plane
designed for the NEW detector. Following this design, the energy
plane of NEXT-100 (see figure~\ref{next100}) will be composed of 60
Hamamatsu R11410-10 photomultiplier tubes located behind the cathode
of the TPC and covering approximately 30\% of its area, as a
compromise between the need to collect as much light as possible and
the need to minimize the number of sensors to reduce cost, technical
complexity and radioactivity. The selected model R11410-10 is a 3''
PMT specially developed for low-background operation, equipped with
a synthetic silica window and a photocathode made of low temperature
bialkali with high quantum efficiency. The PMTs are optically
coupled to sapphire windows using an optical gel with a proper
refractive index. The external face of the windows is coated with
tetraphenyl-butadiene (TPB) to shift the xenon VUV light to blue. In
NEW, the 12~PMTs used are sitting in an unique volume (either vacuum
or nitrogen at 1~bar) separated from the xenon gas volume by a
copper plate. Copper caps, having a thickness similar to that of the
copper plate, are placed behind the PMT bases too.

The structure of the paper is the following. Section~\ref{meas}
summarizes all the measurements performed, describing both the
samples analyzed and the detectors used. Activity results obtained
are collected in section~\ref{resu}, together with the discussion of implications for design and for the NEXT-100 background model.
Finally, conclusions are drawn in section~\ref{disc}.

\section{Measurements}
\label{meas}

The material screening program of the energy readout of the NEXT
experiment is mainly based on germanium gamma-ray spectrometry using
ultra-low background detectors operated at a depth of 2450~m.w.e.,
from the Radiopurity Service of LSC; being a non-destructive
technique, the actual components to be used in the experiment can be
analyzed. Some complementary measurements have been made by GDMS,
performed by Evans Analytical Group in France, providing
concentrations of U, Th and K.
The Radiopurity Service of LSC offers several detectors to measure
ultra-low level radioactivity \cite{gelsc}. They are p-type
closed-end coaxial 2.2-kg High Purity germanium detectors, from
Canberra France, with aluminum or copper cryostats and relative
efficiencies\footnote{Efficiency relative to a $3''\times3''$ NaI
detector at 1332~keV and for a distance of 25~cm between source and
detector.} in the range from 100 to 110\%. Data acquisition is based
on Canberra DSA 1000 modules and shielding consists of 5 or 10 cm of
copper in the inner part surrounded by 20 cm of low activity lead,
flushed with nitrogen gas to avoid airborne radon intrusion. The
measurements related with the energy readout were carried out at LSC
using different detectors (named GeAnayet, GeAspe, GeLatuca,
GeOroel, GeTobazo and Obelix). The detection efficiency required to
quantify activities was estimated by Monte Carlo simulations based
on the Geant4 \cite{geant4} code for each sample, accounting for
intrinsic efficiency, the geometric factor and self-absorption at
the sample; the uncertainty in this efficiency is estimated to be
10\% and it is included in the calculation of the uncertainty for
the derived activities. Activities of different sub-series in the
natural chains of $^{238}$U, $^{232}$Th and $^{235}$U as well as of
common primordial, cosmogenic or anthropogenic radionuclides like
$^{40}$K, $^{60}$Co and $^{137}$Cs have been evaluated by analyzing
the most intense gamma lines of different isotopes; activities have
been quantified when possible and upper limits with a 95.45\% C.L.
have been derived otherwise. More details on the detectors, their
background counting rates, validation of efficiency estimates and
the analysis procedure can be found at \cite{jinstrp,trackingrp}.

Table~\ref{gedata} summarizes the measurements performed for the
samples analyzed in this work, indicating material and supplier, the
detector used, the size of the sample and the live time of data
taking. All the samples were cleaned in an ultrasonic bath and with
pure alcohol before starting the screening, unless this could damage
the component.

\begin{table}
\caption{Information on measurements performed using the germanium
detectors of the LSC Radiopurity Service: component and supplier,
detector used, sample size (mass or number of units) and screening
live time. The corresponding row number of table~2 where the
activity values obtained for each sample are reported is also
quoted. For the measurements of second row for PMT R11410-10,
Hamamatsu, see explanations in section~3.1.}
\begin{tabular}{lcccc}
\\
\hline
Component, Supplier & \# in table~2 &  Detector & Sample size &  Time (d) \\
\hline
PMT R11410-10, Hamamatsu & 1 & GeAnayet & 1 unit & 33.7 \\
PMT R11410-10, Hamamatsu & 2 & GeAnayet & 3 units$\times$18 runs &\\
\hline

Capacitors 1.5~$\mu$F, AVX & 3 & GeLatuca & 392 units (0.16 g/unit) & 37.8 \\
Capacitors 4.7~$\mu$F, AVX & 4 & GeAnayet & 156 units (0.33 g/unit) & 28.0 \\
Polypropylene capacitors, Vishay & 5 & GeAspe & 46 units (8 g/unit) & 22.5 \\
Resistors, Finechem & 6 & GeLatuca & 1200 units (8.8 mg/unit) & 38.5 \\
Resistors, KOA RS & 7 & GeTobazo & 100 units (16.2 mg/unit) & 32.2 \\
Resistors, Mouser & 8 & Obelix & 100 units (9 mg/unit) & 54.1 \\
Pin receptacles, Mill Max & 9 & GeLatuca & 1535 units (51 mg/unit) & 31.9 \\
Thermal epoxy, Electrolube & 10 & GeLatuca & 706 g & 40.4 \\
Epoxy 2011, Araldite & 11 & GeLatuca & 1712 g& 29.6 \\
Solder paste, Multicore & 12 & GeLatuca & 457 g & 44.3 \\
Kapton-Cu cable, Allectra & 13 & GeAspe & 352 g & 12.2 \\
Cuflon, Polyflon & 14 & GeOroel & 1876 g & 24.3 \\
Kapton substrate, Flexible Circuit & 15 & GeAnayet & 50 units (0.61
g/unit) & 54.7\\  \hline

Windows, Precision Sapphire Technologies & 16 & GeAnayet & 527 g & 44.9 \\
Optical gel, Nye Lubricants & 17 & GeAnayet & 53.5 g & 58.3 \\
TPB, Sigma Aldrich & 18 & GeAnayet & 4.1 g & 38.3 \\
PEDOT:PSS, Aldrich Chemistry & 19 & GeAspe & 115 ml & 77.1 \\
M4 screws (manual cleaning) & 23 & GeLatuca & 40 units (2.4 g/unit) & 30.3 \\
M4 screws (Alconox cleaning) & 24 & GeLatuca & 267 units (2.4 g/unit) & 56.6 \\
Vacuum grease, Apiezon M & 25 & GeAspe & 85.4 g& 44.4 \\
Copper CuA1, Lugand Aciers & 26  & GeOroel & 94 kg & 68.6 \\
CuSn braid, RS & 29 & GeAspe & 1875 g & 38.2 \\
 \hline

\end{tabular}
\label{gedata}
\end{table}

\section{Results}
\label{resu}

The activity results obtained for the samples analyzed dealing with
the energy readout plane are all summarized in table~\ref{rpm};
reported errors correspond to $1\sigma$ uncertainties including both
statistical and efficiency uncertainties. In the following, each
sample is described and the corresponding results discussed.

\footnotesize
\begin{landscape}
\begin{longtable}{p{0.2cm}p{3.5cm}p{1.2cm}p{1.2cm}p{1.5cm}p{1.5cm}p{1.5cm}p{1.5cm}p{1cm}p{1.2cm}p{1.2cm}p{1cm}}

\hline
\textbf{\#} & \textbf{Component} &  \textbf{Technique} & \textbf{Unit} & \textbf{$^{238}$U} & \textbf{$^{226}$Ra} & \textbf{$^{232}$Th} &\textbf{$^{228}$Th} & \textbf{$^{235}$U}& \textbf{$^{40}$K}  & \textbf{$^{60}$Co}& \textbf{$^{137}$Cs}\\
 \hline
\endfirsthead
(Continuation)\\
\hline
\textbf{\#} & \textbf{Component} & \textbf{Technique} & \textbf{Unit} & \textbf{$^{238}$U} & \textbf{$^{226}$Ra} &\textbf{$^{232}$Th} &\textbf{$^{228}$Th} & \textbf{$^{235}$U}& \textbf{$^{40}$K}  & \textbf{$^{60}$Co}& \textbf{$^{137}$Cs}\\
\hline
\endhead
&(Follows at next page)\\
\endfoot
\endlastfoot
\hline

1 & PMT & Ge & mBq/PMT & $<$187 & $<$1.8 &  $<$5.4 & $<$3.4 & $<$1.6 & $<$29 & 2.82$\pm$0.27 & $<$0.6 \\
2 & PMT & Ge & mBq/PMT & $<$69 & 0.35$\pm$0.08 & $<$2.1 &0.53$\pm$0.12 & 0.43$\pm$0.11  &  12.1$\pm$1.6 & 3.80$\pm$0.27 &$<$0.3 \\
\hline

3 & Capacitors 1.5~$\mu$F, AVX & Ge&  $\mu$Bq/unit&   $<$360&    72$\pm$3    &49$\pm$3& 38$\pm$2&&      71$\pm$9&   $<$1&  $<$1\\
4 & Capacitors 4.7~$\mu$F, AVX & Ge&  $\mu$Bq/unit& $<$900&    123$\pm$7    &95$\pm$7& 86$\pm$6&&     123$\pm$21&   $<$3&  $<$2\\
5 & Capacitors Vishay & Ge & mBq/unit & 10.4$\pm$2.7 & 5.29$\pm$0.25 & 8.52$\pm$0.51 & 8.75$\pm$0.49 & & 5.29$\pm$0.57 & $<$0.036 & $<$0.043 \\
6 & Resistors Finechem &Ge&    $\mu$Bq/unit&   85$\pm$23&  4.1$\pm$0.3 &5.6$\pm$0.5&   4.4$\pm$0.3 &&  83.6$\pm$8.7&   $<$ 0.2 &104$\pm$11\\
7 & Resistors KOA RS & Ge&  $\mu$Bq/unit& $<$852 & $<$7.7 & $<$14 & $<$4.1 & $<$3.5 & $<$29 & $<$2.1 & $<$1.5  \\
8 & Resistors Mouser & Ge & $\mu$Bq/unit& $<$182 & $<$7.0 & 5.3$\pm$1.5 & $<$8.0 & 3.7$\pm$1.1 & $<$37 & $<$1.7 & $<$1.8 \\

9 & Pin receptacles & Ge& $\mu$Bq/unit&   217$\pm$42  &$<$1.1    &5.6$\pm$0.5    &4.5$\pm$0.4&   6.1$\pm$0.5 &20.5$\pm$2.4   &$<$0.3    &$<$0.2\\
10 & Thermal epoxy& Ge& mBq/kg&(1.0$\pm$0.2)10$^{3}$ &169.4$\pm$7.9& 52.1$\pm$3.7 &54.4$\pm$3.2   &&  105$\pm$12& $<$1.1& $<$1.3\\
11 & Epoxy Araldite & Ge & mBq/kg & $<$182 & $<$1.4 & $<$3.7 & $<$2.5 & $<$0.8 & 15.0$\pm$2.4 & $<$0.4 &  $<$0.4 \\
12 & Solder paste & Ge & mBq/kg & $<$310 & $<$2.7 & $<$4.7 & $<$2.5 & $<$5.2 & $<$13 & $<$1.0 & $<$1.6 \\
13 & Kapton-Cu cable & Ge & mBq/kg & $<$1.1$\times10^{3}$ & 46.8$\pm$3.3 & $<$40 & $<$32 & & 166$\pm$27 & $<$5.2 & $<$4.4 \\
14 & Cuflon & Ge & mBq/kg & $<$33 & $<$1.3 & $<$1.1 & $<$1.1 & $<$0.6 & 4.8$\pm$1.1 & $<$0.3 & $<$0.3\\
15 & Kapton substrate & Ge & $\mu$Bq/unit& $<$2.8$\times10^{3}$ & $<$23 & 77$\pm$13 & 43.9$\pm$7.2 & $<$18 & $<$216 & $<$6.4 & $<$6.7 \\
\hline

16 & Sapphire windows & Ge & mBq/kg & $<$275 &  $<$ 2.7 &  $<$7.6 & $<$5.5 &  $<$2.1 &  $<$18 & $<$0.7&  $<$1.0  \\
17 & Optical gel & Ge & mBq/kg & $<$1.7$\times10^{3}$ & $<$22 & $<$49 & $<$18 & $<$16 & $<$173 & $<$4.5 & $<$5.8 \\
18 & TPB & Ge & Bq/kg & $<$23 & $<$0.17 & $<$0.57 & $<$0.15 & $<$0.11 & $<$1.7 & $<$0.05 & $<$0.05 \\
19 & PEDOT:PSS & Ge & $\mu$Bq/ml & $<$626 & $<$6.9 & $<$23 & $<$4.8 & $<$3.9 & 49$\pm$11 & $<$1.7 & $<$1.8 \\
20 & Brazing paste &GDMS & $\mu$Bq/kg & 55$\pm$10 && 49$\pm$4 &&& $<$31 & & \\
21 & Brass bolts &GDMS & $\mu$Bq/kg & 8.9$\pm$0.7 && 6.9$\pm$0.2 & & &  $<$31 & & \\
22 & SS screws &GDMS & mBq/kg & 3.25$\pm$0.25 & & 0.57$\pm$0.08 & && $<$0.19 & & \\
23 & M4 screws, manual clean & Ge & $\mu$Bq/unit & $<$2.2$\times10^{3}$ & $<$21 & $<$60 & 20.0$\pm$4.6 & $<$12 & $<$93 & 14.0$\pm$1.8 & $<$6.0 \\
24 & M4 screws, Alconox clean & Ge & $\mu$Bq/unit & $<$616 & $<$8.6 & 14.9$\pm$3.4 & 17.4$\pm$1.8 & 3.7$\pm$1.0 & $<$19 & 13.4$\pm$1.1 & $<$1.4 \\
25 & Vacuum grease & Ge & mBq/kg & < 1.0$\times10^{3}$ & $<$10 & $<$ 43 & $<$8.5 &  $<$6.1 & $<$49 & $<$3.5 & $<$2.9  \\
26 & CuA1 & Ge& mBq/kg& $<$4.1&    $<$0.16&   $<$0.15&   $<$0.13&   $<$0.17    &$<$0.37&  0.04$\pm$0.01   &$<$0.04\\
27 & CuA1 & GDMS&   $\mu$Bq/kg& $<$12&&     $<$4.1   &&&     62   &&  \\
28 & CuC1 & GDMS&  $\mu$Bq/kg& 25$\pm$5 &&15$\pm$4 &&& 190 &&  \\
29 & CuSn braid & Ge & mBq/kg & $<$168 & $<$2.4 & $<$7.1 & $<$2.1 &$<$1.8 & $<$14 & $<$0.6 & $<$0.5\\

\hline

\caption{Activities measured for the energy readout components used
in NEXT. Results reported for $^{238}$U and $^{232}$Th correspond to
the upper part of the chains and those of $^{226}$Ra and $^{228}$Th
give activities of the lower parts. For component \#2, see
explanations in section~3.1.}
\label{rpm} 
\end{longtable}
\end{landscape}
\normalsize

\subsection{Photomultipliers}
\label{pmtsec}

The photomultiplier tubes are the basic element of the energy
readout plane of NEXT. The number of required PMTs is 12 for NEW and
60 for NEXT-100. All the 55 available units of the selected model,
Hamamatsu\footnote{http://www.hamamatsu.com} R11410-10, have been
screened using the same germanium detector (GeAnayet) at LSC during
2013, 2014 and 2015. First, a single photomultiplier was analyzed
(row \#1 of table~\ref{rpm}) and then eighteen runs were carried
out, each one with three different units placed around the detector
on a teflon support; data taking at each run ranged from 18.3 to
41.0~days per run and several reference backgrounds were measured in
between. For the detection efficiency simulation, emissions are
assumed to be uniformly generated in the kovar PMT enclosure.
Activities for the three PMTs at each independent measurement were
deduced and are summarized in table~\ref{PMTsresults}. The results
obtained from different runs are roughly compatible; $^{60}$Co
activity has been always quantified, those of $^{40}$K and $^{54}$Mn
only in the most sensitive runs while upper limits have been
obtained in general for all the other common isotopes.

\begin{table}
\caption{Activities (in mBq) measured at each individual run
screening altogether three different PMTs per run. Results reported
for $^{238}$U and $^{232}$Th correspond to the upper part of the
chains and those of $^{226}$Ra and $^{228}$Th give activities of the
lower parts.}
\begin{tabular}{lccccccccccc}
\\
\hline

Run & Time & $^{238}$U & $^{226}$Ra & $^{232}$Th & $^{228}$Th &
$^{235}$U & $^{40}$K  & $^{60}$Co & $^{137}$Cs & $^{54}$Mn\\
& (d) & & & & & & & & & \\ \hline

1 & 30.6 & $<$266 & $<$2.7 & $<$7.8 & $<$6.1 & $<$3.4 & 30.4$\pm$8.1 &  12.2$\pm$1.0 & $<$1.0 & 1.1$\pm$0.3 \\

2 & 20.2 & $<$340 & $<$3.3 & $<$9.1 & $<$7.9 & $<$4.4 & $<$60 & 11.1$\pm$0.9 & $<$1.3 & $<$1.7 \\

3 & 22.0 & $<$320 & $<$3.3& $<$10& $<$7.3& $<$3.8 & 39.5$\pm$9.4 & 10.4$\pm$0.9& $<$1.2 & $<$1.3 \\

4 & 18.3 & $<$351 & $<$4.3 & $<$9.5& $<$9.6 &$<$3.8 &41$\pm$10 & 11.5$\pm$1.0 & $<$1.4 & 1.1$\pm$0.3 \\

5 & 41.0 & $<$229 & $<$2.5 & $<$6.6 & $<$7.2 &$<$3.1 & 32.3$\pm$7.7 & 10.8$\pm$0.8 & $<$0.8 & 1.1$\pm$0.2 \\

6 & 39.8 & $<$232 & $<$2.7 & $<$9.3 & $<$6.2 & $<$3.1 & 30.0$\pm$7.5 & 10.4$\pm$0.8 & $<$0.8 & 1.0$\pm$0.3 \\

7 & 23.0 & $<$317 & $<$4.3 & $<$9.1 & $<$6.5 & $<$2.6 & 34.2$\pm$8.9 &  10.9$\pm$0.9 & $<$1.1 & $<$1.4 \\

8 & 38.7 & $<$234 & $<$3.0 & $<$9.3 & $<$5.9 & $<$3.2 & 35.5$\pm$7.9 &  12.3$\pm$1.0 & $<$0.8 & $<$1.2 \\

9 & 19.5 & $<$330 & $<$3.4 & $<$9.2 & $<$10 & $<$4.3 & 34.8$\pm$9.5
& 11.3$\pm$0.9 & $<$1.2 & $<$1.8 \\

10 & 27.8 & $<$293 & $<$3.9 & $<$9.4 & $<$6.3 & $<$4.4 & 36.9$\pm$8.6 & 12.0$\pm$0.9 & $<$1.0 & $<$1.4 \\

11 & 30.4 & $<$286 & 2.7 $\pm$0.8 & $<$11 & $<$6.8 & $<$2.2 & 38.0$\pm$8.7 & 11.1$\pm$0.9 & $<$1.3 & $<$1.4 \\

12 & 33.5 & $<$405 & $<$3.2 & $<$6.9 & $<$6.5 & $<$3.8 & 28.3$\pm$7.9 & 11.0$\pm$0.9 & $<$0.8 & 0.9$\pm$0.3 \\

13 & 23.6 & $<$361 & $<$5.2 & $<$8.3 & $<$8.7 & 2.5$\pm$0.7 & 50$\pm$10 & 11.4$\pm$0.9 & $<$1.1 & 1.2$\pm$0.3 \\

14 & 38.6 & $<$337 & $<$3.7 & $<$8.9 & $<$6.9 & $<$3.0 & 30.7$\pm$7.8 & 11.7$\pm$0.9 & $<$0.8 & 0.9$\pm$0.2 \\

15 & 24.2 & $<$470 & 3.0 $\pm$0.9 & $<$8.6 & $<$6.2 & $<$4.3 & 34.5$\pm$9.0 & 10.3$\pm$0.8 & $<$1.2 & $<$1.4 \\

16 & 24.1 & $<$418 & $<$4.5 & $<$11 & $<$7.9 & $<$4.1 & 41.6$\pm$9.4 & 12.2$\pm$1.0 & $<$1.1 & 0.8$\pm$0.2 \\

17 & 19.3 & $<$443 & $<$4.9 & $<$10 & $<$8.6 & $<$3.5 & $<$61 & 12.1$\pm$1.0 &  $<$1.2 & $<$1.7 \\

18 & 23.6 & $<$448 & $<$5.6 & $<$8.8 & $<$9.0 & $<$2.6 &
45.3$\pm$9.3 & 12.5$\pm$1.0 & $<$1.0 & $<$1.1 \\ \hline

All & 498.2 & $<$83 & 1.05$\pm$0.24 & $<$2.8 & 1.58$\pm$0.35 & 1.30$\pm$0.32 & 36.3$\pm$4.8 & 11.40$\pm$0.81 & $<$0.35 & \\

\hline
\end{tabular}
 \label{PMTsresults}
\end{table}

Since, following these results, it seemed that the activity levels
of all the screened PMTs were similar, a joint analysis of the
available data was attempted in order to further increase the
sensitivity by combining data from the eighteen independent runs
performed with three PMTs altogether corresponding to a total
exposure of 498.2~days. The obtained results are shown in last row
of table~\ref{PMTsresults}. Since it was assumed that all the PMTs
were equivalent, activity values per PMT were estimated just
considering one third of the net signal measured; these results are
shown in row \#2 of table~\ref{rpm}.
Thanks to this joint analysis, the activity of $^{235}$U has been
properly evaluated. Although there is also a clear net signal from
$^{54}$Mn, a direct quantification of the activity has not been
attempted since its half-life (312.3~days) is comparable to the time
span of the measurements and then its decay should be properly taken
into consideration. Concerning the lower parts of the $^{232}$Th and
$^{238}$U chains, several lines show an excess of events above
background statistically significant thanks to the accumulation of
data; therefore, it has been possible to quantify the activity of
some of their isotopes and a robust estimate of the average activity
per PMT of $^{226}$Ra and $^{228}$Th has been achieved.

The same model of PMT has been screened for other experiments
\cite{APR11,luxpmt,pandax} and our results are in very good
agreement with those found by XENON; in particular, $^{60}$Co and
$^{40}$K activities are virtually the same.
\begin{itemize}
\item The XENON1t collaboration has carried out a deep study of the
radioactivity of the new PMT version Hamamatsu R11410-21
\cite{pmtxenon1t}, based on analysis using GDMS and germanium
spectrometry of individual components and units
\cite{xenon1t,rpxenon1t}. The main differences with respect to the
version R11410-10 are the use of Co-free kovar body and high purity
(instead of standard purity) Al seal. Comparing results for the two
versions, it can be concluded that $^{40}$K and $^{235}$U activities
are comparable, $^{60}$Co has been reduced about a factor~4-5 in the
new version and for the lower parts of the natural chains of
$^{232}$Th and $^{238}$U, activity is at the same level, about
one-half mBq/PMT.
\item Results for the version R11410-20 have been presented by the
LUX-ZEPLIN collaboration \cite{lz}. Intrinsic radioactivity of the
component materials to be used in the manufacture has been performed
using several germanium detectors and upper limits set for the total
activity of all components are comparable to the best quantified
activities for other versions; only the upper limit set for $^{40}$K
is a factor $\sim$4 lower. The screening of final tubes is underway.
\end{itemize}

\subsection{PMT bases}
Each PMT base in the NEW set-up is composed of a total of
19~resistors of different electrical resistance, 7~capacitors (5
having a capacitance of 1.5~$\mu$F and 2 with 4.7~$\mu$F) and 18~pin
receptacles fixed on a kapton substrate using epoxy, a copper cap
with a mass of $\sim$50~g and a 1-m-long cable made of kapton and
copper. A paste was used for soldering. All these components used in
the PMT bases have been separately screened; the PMT base design
reflects a careful compromise between performance of electronics
response (particularly concerning charge linearity) and radio
purity.

Base capacitors are Tantalum Solid Electrolytic Chip Capacitors with
Conductive Polymer Electrode, TCJ Series, supplied by
AVX\footnote{http://www.avx.com}. Two samples of units with the
different capacitance and different size and mass have been
screened.
It was possible to quantify the activities from $^{40}$K, $^{232}$Th
and the lower part of the $^{238}$U chain (rows \#3 and 4 of
table~\ref{rpm}). The measured activities in the larger capacitors
are roughly a factor of 2 higher than in the smaller ones, which is
also the ratio between the masses of each unit. The presence of
$^{182}$Ta (beta emitter with Q=1814.3~keV, T$_{1/2}$=114.6~days,
produced by neutron activation on $^{181}$Ta) was identified by
means of several of its gamma emissions. In addition, a sample of
capacitors from Vishay\footnote{http://www.vishay.com} (Metallized
Polypropylene Film Capacitors, 5~$\mu$F,
32$\times$11$\times$21~mm$^{3}$ each unit) having the dielectric
made of polypropylene was screened too. All common radioisotopes
were quantified with activities of a few mBq/unit (row \#5 of
table~\ref{rpm}), which are unacceptable for NEXT, and consequently
the use of these polypropylene capacitors was discontinued.

SMD resistors to be used at the voltage divider from several
suppliers have been screened:
\begin{itemize}
\item SM2 resistors supplied by the Japanese
company Finechem\footnote{http://www.jfine.co.jp} have an alumina
ceramic substrate. The dimensions of each unit are
3.2$\times$1.6$\times$0.55~mm$^{3}$. Activities have been derived
for $^{40}$K as well as for the $^{232}$Th and $^{238}$U chains (row
\#6 of table~\ref{rpm}). In addition, the resistors showed important
activities from $^{134}$Cs and $^{137}$Cs, which could be related to
the Fukushima accident. For $^{134}$Cs (beta emitter with
Q=2058.98~keV, T$_{1/2}$=2.06~y), activity was
32.7$\pm$1.6~$\mu$Bq/unit. The results obtained can be compared with
the ones for SM5D Finechem resistors, showing no Cs activity,
presented at \cite{jinstrp,APR11}; results are roughly consistent
taking into account that the volume of SM5 resistors is four times
the one of SM2 resistors.
\item Another sample consisted of resistors produced by KOA Speer\footnote{http://www.koaspeer.com} and supplied by RS (Thin Film 1206, 62~$\Omega$).
Dimensions of each unit are 3.2$\times$1.6$\times$0.6~mm$^{3}$. None
of the common radioisotopes has been quantified and upper limits to
their activities have been set (row \#7 of table~\ref{rpm}).
\item Finally, resistors from Mouser\footnote{http://www.mouser.com} (62~$\Omega$) were analyzed
too. In this case, activity of some isotopes has been quantified and
upper limits for the other ones have been set (row \#8 of
table~\ref{rpm}).
\end{itemize}
Comparing the results from the three considered resistors, it can be
concluded that the quantified activities or upper limits are at
similar levels of a few $\mu$Bq/unit for all of them; finally,
14~units from Finechem and 5 from RS have been selected.

Pin receptacles from Mill Max\footnote{http://www.mill-max.com}
(model 0327-0-15-15-34-27-10-0) having a shell made of brass alloy
360 were screened. Activities for most of the common radioisotopes
have been quantified (row \#9 of table~\ref{rpm}).

Thermally conductive epoxy produced by
Electrolube\footnote{http://www.electrolube.com} (division of HK
WENTWORTH LTD) was screened to be used on the PMT bases to dissipate
heat. To prepare the sample the epoxy (EER2074A) and corresponding
hardener (EER2074B) were mixed inside a clean container following
specifications. A sample of the epoxy Araldite
2011\footnote{http://www.go-araldite.com} was screened too. A
mixture of Araldite2011-A (resin) and Araldite 2011-B (hardener) was
prepared. As shown in rows \#10-11 of table~\ref{rpm}, Araldite
epoxy showed a better radiopurity since only $^{40}$K was quantified
and upper limits for the other common radioisotopes were derived;
these are more stringent than those available from \cite{busto}.
Therefore, the epoxy Araldite 2011 was finally used in the PMT
bases.

A sample of lead-free SnAgCu solder paste supplied by Multicore
(ref. 698840) was screened and results are presented in row \#12 of
table~\ref{rpm}. $^{108m}$Ag, induced by neutron interactions and
having a half-life of T$_{1/2}$=438~y, has been identified in the
paste, with an activity of (5.26$\pm$0.40)~mBq/kg, while upper
limits of a few mBq/kg have been set for the common radioactive
isotopes.

A roll of the kapton-copper cable supplied by Allectra
company\footnote{http://www.allectra.com} was screened. Activities
of some isotopes were quantified (row \#13 of table~\ref{rpm}) and
in addition, the presence of $^{108m}$Ag can be reported through the
identification of its most intense gamma lines.

Concerning the base substrate, cuflon and kapton have been
considered. Cuflon$\circledR$ offers low activity levels, as shown
in the measurement of samples from Crane
Polyflon\footnote{http://www.polyflon.com} by GERDA \cite{BUD09} and
at \cite{NIS09}, using both ICPMS and Ge gamma spectroscopy. As
presented in \cite{jinstrp}, a measurement of Polyflon cuflon made
of a 3.18-mm-thick PTFE layer sandwiched by two 35-$\mu$m-thick
copper sheets was made for NEXT and results are shown in row \#14 of
table~\ref{rpm}. Only activity of $^{40}$K could be quantified.
Although cuflon could have been used too, kapton was finally
selected and a sample of the produced 0.5-mm-thick base substrates
by Flexible Circuit\footnote{http://www.flexiblecircuit.com} was
screened. Upper limits were set for all common radioisotopes except
for $^{232}$Th (row \#15 of table~\ref{rpm}).

\subsection{Windows, PMT enclosures and other components}

Other components also used in the energy readout plane have been
taken into consideration. Four sapphire crystals to be used as PMT
windows were screened; each crystal was 6~mm thick with a diameter
of 83.8~mm. They were measured on a teflon support for protection.
No isotope was quantified (row \#16 of table~\ref{rpm}).
Since the upper limits obtained from germanium spectrometry are
quite high, results from Neutron Activation Analysis (NAA) presented
at measurement \#155 in \cite{LEO08} by the EXO collaboration have
been considered for the moment in the development of the NEXT-100
background model.

The silicone-based optical gel from Nye Lubricants
Inc.\footnote{http://www.nyelubricants.com} (SmartGel NyoGel
OCK-451), used for PMT coupling, was screened. The optical fluid and
thickening agent were mixed in the clean room of LSC and left there
for 26~hours in order to form a solid disk. The sample was prepared
on a clean container. No isotope was quantified and upper limits
were set for all of them (row \#17 of table~\ref{rpm}). The quantity
to be used per window is estimated to be about 2~g.

A sample of the TPB material coating the enclosure windows, supplied
by Sigma Aldrich\footnote{http://www.sigmaaldrich.com}, was
analyzed. The powder was prepared inside a clean Petri dish. Upper
limits on the specific activity were set for all the radioisotopes
(row \#18 of table~\ref{rpm}). Due to the small mass of our sample,
better results for this material from the same supplier can be found
at \cite{SNO,TPBGS}; activities for $^{238}$U and $^{232}$Th at the
level of tenths of mBq/kg or even lower are reported there.

A sample of PEDOT:PSS (1.3 wt\% dispersion in water) also from
Aldrich Chemistry to be used as conductive coating on sapphire
windows was screened too (row \#19 of table~\ref{rpm}); only the
activity of $^{40}$K could be quantified. It is applied by
spin-coating and then dried to evaporate water, resulting in a
$\sim$100-nm-thick layer.

Other materials or components to be used at the PMT enclosures were
analyzed by GDMS. A sample of brazing paste made of 72\% Ag and 28\%
Cu with dimensions 12$\times$12$\times$12~mm$^{3}$ was measured
quantifying the U and Th content (row \#20 of table~\ref{rpm}). M4
vented screws made of 316 stainless steel were screened; the mass of
each 2-cm-long unit is 2.32~g. A sample of a M4 bolt made of brass,
with length 22.65~mm and mass 3.08~g, was also analyzed and the
results are shown in rows \#21-22 of table~\ref{rpm}. Since 28 units
are needed per PMT can, brass bolts were preferred instead of the
vented screws from the radiopurity point of view.

However, in principle, stainless steel has been used for mechanical
reasons in the NEW set-up. Samples of M4 screws were screened using
germanium detectors. Since these screws were pre-greased, a cleaning
procedure was necessary to remove the grease, which could affect the
purity of the xenon gas and is expected to be non-radiopure; two
options were analyzed. A manual cleaning was applied to a sample, by
wiping the screws several times with alcohol by hand, cleaning them
in an ultrasound bath with soap and afterwards rinsing with alcohol.
For another sample of screws, cleaning was made using
Alconox\footnote{http://alconox.com} detergent 8 (5\% solution in
water) in an ultrasound bath. Activities from $^{60}$Co, $^{235}$U
and for the $^{232}$Th chain have been quantified (rows \#23-24 of
table~\ref{rpm}). Results obtained for the two samples with
different cleaning procedures are compatible; upper limits derived
for the sample cleaned using Alconox are more stringent thanks to
the larger number of screened units. Once the original grease on the
screws is removed, vacuum grease must be used; a sample of
Apiezon\footnote{http://www.apiezon.com} M grease designed for high
vacuum applications was analyzed and only upper limits were set
(rows \#25 of table~\ref{rpm}). Due to the relatively large
contribution of these M4 screws (quoted for reference in
table~\ref{comcont}) the use of the much more radiopure brass bolts
is foreseen in the NEXT-100 detector.

Two types of copper supplied by Lugand Aciers
company\footnote{http://lugand-aciers.fr} were screened for use at
the energy plane: CuA1\footnote{This type of copper is also referred
as Cu-ETP (Electrolytic Tough Pitch) or C11000. Its copper purity is
99.90\% (minimum).} for PMT enclosures and base caps and
CuC1\footnote{This type of copper is also referred as Cu-OF
(Oxygen-Free) or C10200. Its copper purity is 99.95\%.} for both the
tracking and energy readout plates. The weight of each PMT copper
enclosure is 4.1~kg and that of the energy plate 475.6~kg. A large
mass of CuA1 copper was accumulated to carry out a measurement using
a germanium detector at LSC; a special cleaning procedure typically
used for copper shieldings was performed at LSC before the
measurement, consisting of soap cleaning, nitric acid etching,
passivation with citric acid and drying. As shown in row \#26 of
table~\ref{rpm}, only  $^{60}$Co activity was quantified. Peaks from
other cobalt isotopes (also common cosmogenic products in copper
induced by the exposure of the material to cosmic nucleons at sea
level) were identified ($^{56}$Co with T$_{1/2}$=77.27~d and
$^{58}$Co with T$_{1/2}$=70.86~d); since their half-lives are of the
order of the live time in the screening measurement, the direct
quantification of their activities was not performed. GDMS analysis
was additionally made for two samples, with dimensions
12$\times$12$\times$12~mm$^{3}$, made of CuA1 and CuC1 copper,
having received the same cleaning protocol. Following results
presented in rows \#27-28 of table~\ref{rpm}, GDMS upper limits for
the CuA1 sample are much lower than the ones derived from germanium
spectrometry in \#26. The good results obtained for this copper led
to it being used also for the shielding against gamma radiation to
be placed inside the pressure vessel made of 316Ti stainless steel.
Results for CuA1 from Lugand Aciers, at the level of a few
$\mu$Bq/kg for $^{238}$U and  $^{232}$Th, are equivalent to those
obtained for C10100 copper supplied by the Luvata company
\cite{jinstrp} and similar to those for the Norddeutsche
Affinerie\footnote{Re-branded as Aurubis, http://www.aurubis.com}
copper \cite{ARI04}. Results at or even below tenths of $\mu$Bq/kg
have been presented for electroformed and also commercial copper
analyzed by ICPMS for the Majorana experiment \cite{pnnl,majorana}.

A sample of the CuSn braid used to dissipate heat at the PMT cans
was measured using a germanium detector. It is soft tinned copper
wire braid 2536P provided by RS. The total length of the sample was
19.4~m. Upper limits were set for all the common radioisotopes (row
\#29 of table~\ref{rpm}).

Finally, the expected contribution to the background level in the
region of interest of NEXT-100, assuming a NEW-like design with
60~PMT modules, from the activities of all the relevant components
of the energy plane has been evaluated by Monte Carlo simulation
using the Geant4 package (see details at \cite{nextsensitivity}) and
is reported in table~\ref{comcont}. A first estimate of the
contribution of the energy plane was already made in
\cite{nextsensitivity}, considering the main  components (PMTs, PMT
enclosures and sapphire windows). Using the upper limits or the
quantified activity of $^{208}$Tl and $^{214}$Bi presented here for
all the selected components for the energy readout, the expected
rate in the region of interest for the neutrinoless double beta
decay of $^{136}$Xe is below $2.4\times10^{-4}$~counts
keV$^{-1}$~kg$^{-1}$~y$^{-1}$, complying with the requirements to
achieve the desired sensitivity. As it can be concluded from
table~\ref{comcont}, PMTs and base capacitors are the dominant
contributors accounting for 29.9\% and 36.5\%, respectively, of the
total estimated background.

\begin{table}
\caption{Summary of the expected background rate in the region of
interest for NEXT-100 assuming a NEW-like design from the most
relevant components of the energy readout plane, estimated by Monte
Carlo simulation. Activities considered are indicated in the second
column, either from some references or from entries in table~2.
Upper limits are at 95.5\%~C.L. All numbers correspond to
10$^{-4}$~counts keV$^{-1}$~kg$^{-1}$~y$^{-1}$. Contribution marked
with (*) is not included in the total sum (see text). As a
reference, the estimate of the background rate from the energy plane
in \cite{nextsensitivity} considering just PMTs, PMT enclosures and
sapphire windows was $1.0\times10^{-4}$~counts
keV$^{-1}$~kg$^{-1}$~y$^{-1}$.}
\begin{tabular}{lcccc}
\\
\hline Component& Activity & Rate from $^{208}$Tl & Rate from
$^{214}$Bi & \\ \hline

PMTs & \#2 & 0.35$\pm$0.08 & 0.37$\pm$0.08 \\
Base capacitors 1.5~$\mu$F &   \#3   & 0.125$\pm$0.007 & 0.377$\pm$0.016    \\
Base capacitors 4.7~$\mu$F &   \#4   & 0.113$\pm$0.008 & 0.258$\pm$0.015    \\
Finechem base resistors  &   \#6   &       0.040$\pm$0.003   &       0.060$\pm$0.004   \\
RS base resistors &  \#7 & $<$0.013 & $<$0.040 \\
Pin receptacles at base &   \#9   &       0.053$\pm$0.005   &   $<$0.02        \\
Epoxy Araldite & \#11 & $<$0.014 & $<$0.012 \\
Kapton-Cu cable & \#13 & $<$0.006 & 0.010$\pm$0.001 \\
Kapton substrate &  \#15 & 0.036$\pm$0.006 & $<$0.030 \\
Copper caps at base &   \#27  &   $<$1  10$^{-4}$    &   $<$6 10$^{-4}$    \\
Sapphire windows    &   Ref. \cite{LEO08}  &       0.018$\pm$0.004   &   $<$0.11        \\
Optical gel &   \#17  &   $<$0.041        &   $<$0.13        \\
TPB & Ref. \cite{TPBGS} & (1.3$\pm$0.3) 10$^{-5}$ & (1.1$\pm$0.4) 10$^{-4}$  \\
PEDOT:PSS & \#19 & $<$2.6 10$^{-6}$ & $<$9.4 10$^{-6}$ \\
Brazing paste & \#20 & (2.1$\pm$0.2) 10$^{-4}$ & (6.0$\pm$1.1) 10$^{-4}$  \\
M4 screws (*) &   \#24  &       0.52$\pm$0.05   &    $<$0.66       \\
Brass bolts &  \#21  & (6.4$\pm$0.2) 10$^{-4}$ & (2.1$\pm$0.2) 10$^{-3}$ \\
PMT copper enclosures   &   \#27  &   $<$0.001        &   $<$0.002        \\
Copper plate    &   \#28  &       0.054$\pm$0.014    &       0.065$\pm$0.013    \\
CuSn braid & \#29 & $<$0.014 & $<$0.025 \\
\hline

Total & & $<$0.88 & $<$1.51 \\ \hline

\end{tabular}
\label{comcont}
\end{table}

Apart from the bulk emissions from the measured activity,
radon-induced background from surface deposition or emanated radon,
as an intermediate decay product of the uranium and thorium series,
can be a concern when requiring ultra-low background conditions.
Radon can emanate from detector components and be transported to the
active volume through the gas circulation. The progeny of radon is
positively charged and adhere to surfaces or dust particles; it
drifts toward the TPC cathode and the subsequent $^{214}$Bi and
$^{208}$Tl decays are a potential background source. Radon
contamination in the xenon gas causes two different types of
background events: $\beta$ tracks from the decay of $^{214}$Bi in
the active volume, and photoelectrons generated by gamma rays
emitted, for the most part, from the TPC cathode. This background
source was carefully analyzed in \cite{nextsensitivity}, evaluating
the corresponding background rate generated in NEXT-100 in terms of
the activity of $^{222}$Rn; it was concluded that in order for this
background to contribute, at most, at the level of
10$^{-5}$~keV$^{-1}$~kg$^{-1}$~y$^{-1}$, radon activities in the
xenon gas below a few mBq per cubic meter would be required. The
design of NEXT-100 minimizes the use of materials and components
known to emanate radon in high rates, such as plastics, cables or
certain seals and cleaning of surfaces close to the active volume is
foreseen. In addition, radon emanation measurements are being
carried out for different components used, including those of the
energy readout plane, in collaboration with the Jagiellonian
University (Cracow, Poland) using a cryogenic radon detector;
preliminary results point to acceptable activities of $^{220}$Rn and
$^{222}$Rn for the PMTs, the TPB-coated PMT windows and the
kapton-copper cable.

\section{Conclusion}
\label{disc}

A thorough control of the material radiopurity is being performed in
the construction of the NEXT double beta decay experiment to be
operated at LSC, mainly based on activity measurements using
ultra-low background gamma-ray spectrometry with germanium detectors
of the Radiopurity Service of LSC and complementary GDMS analysis.
Radiopurity information is helpful not only for the selection of
sufficiently radiopure materials, but also for the development of
the detector background model in combination with Monte Carlo
simulations. Many of the components to be actually used in the
experiment, like the PMT units, have been directly screened.

The design of a radiopure energy readout plane for the NEXT
detection system, which must be in direct contact with the gas
detector medium, was a challenge (as it was for the tracking plane
\cite{trackingrp}) since photomultipliers and electronic components
can typically have much higher activity levels than those tolerated
in ultra-low background experiments. Selection of in-vessel
components was performed in parallel to its design.

Photomultiplier tubes are the main element of the energy readout in
the NEXT detector. All the available units of the selected PMT
model, Hamamatsu R11410-10, were screened in 3-unit groups using the
same germanium detector at LSC. Compatible activities were
registered for all runs and a joint analysis of all the accumulated
data allowed us to quantify average activities of not only $^{60}$Co
and $^{40}$K but also of the isotopes in the lower part of the
$^{238}$U and $^{232}$Th chains, of uppermost relevance for NEXT-100
background. The found activities are similar to those measured for
the new version of the PMT R11410-21 \cite{xenon1t}, except for
$^{60}$Co, which is about five times larger in the version
considered here. In addition, most of the components accompanying
PMTs at their bases and enclosures were analyzed, including sapphire
windows, optical gel, capacitors, resistors, cables, epoxy, bolts,
screws and copper; results are summarized in table~\ref{rpm}. The
procurement of large samples with a huge number of pieces made it
possible to measure activities at the level of $\mu$Bq/unit for
several components. Some items were disregarded due to bad
radiopurity and replaced by other ones.

The construction of a precise NEXT-100 background model is based on
Geant4 simulation and it allows us to evaluate the experimental
sensitivity. After the first estimate made in
\cite{nextsensitivity}, the contribution from the energy plane to
the background level in the region of interest for the neutrinoless
double beta decay of $^{136}$Xe has been reanalyzed here considering
all the material radiopurity information collected and assuming a
NEW-like design; as shown in table~\ref{comcont}, PMTs and base
capacitors are found to be the dominant contributors. The expected
rate is below $2.4\times10^{-4}$~counts
keV$^{-1}$~kg$^{-1}$~y$^{-1}$, satisfying the requirements to
achieve the desired sensitivity. But this contribution could be
further reduced thanks to some changes implemented in the PMT bases
design or other components for the final NEXT-100 detector.

\acknowledgments Special thanks are due to LSC directorate and staff
for their strong support for performing the measurements at the LSC
Radiopurity Service. We are really grateful to Grzegorz Zuzel for
the radon emanation measurements. The NEXT Collaboration
acknowledges support from the following agencies and institutions:
the European Research Council (ERC) under the Advanced Grant
339787-NEXT; the Ministerio de Econom\'ia y Competitividad of Spain
under grants FIS2014-53371-C04 and the Severo Ochoa Program
SEV-2014-0398; the GVA of Spain under grant PROMETEO/2016/120; the
Portuguese FCT and FEDER through the program COMPETE, project
PTDC/FIS/103860/2008; the U.S.\ Department of Energy under contracts
number DE-AC02-07CH11359 (Fermi National Accelerator Laboratory) and
DE-FG02-13ER42020 (Texas A\&M); and the University of Texas at
Arlington.

\end{document}